\documentclass[11pt,dvips]{article}

\usepackage{epsfig,times}
\usepackage{picinpar}
\usepackage{wrapfig}
\usepackage{floatflt}
 \makeatletter
\renewcommand{\section}{\@startsection{section}{1}{0in}
    {0.4\baselineskip}{0.1\baselineskip}{\Large\bf}}
\renewcommand{\subsection}{\@startsection{subsection}{2}{0in}
    {0.25\baselineskip}{-\baselineskip}{\large\bf}}
\renewcommand{\subsubsection}{\@startsection{subsubsection}{3}{0in}
    {0.1\baselineskip}{-\baselineskip}{\normalsize\bf}}
\makeatother

\begin{document}
\makeatletter\newcommand{\ps@icrc}{
\renewcommand{\@oddhead}{\slshape{HE.6.1.10}\hfil}}
\makeatother\thispagestyle{icrc}
\begin{center}

{\LARGE \bf Horizontal Tau air showers from mountains in deep
valley:\\ Traces of UHECR neutrino tau }
\end{center}

\begin{center}
{\bf D. Fargion $^1$, A. Aiello $^2$, R. Conversano}\\ {\it $^{1}$
Physics Department, Rome University 1, and INFN, Rome1, P.za Aldo
Moro 2 Rome, ITALY \\ $^{2}$ Physics Department, Rome University
1}
\end{center}

\begin{center}
{\large \bf Abstract\\}
\end{center}
\vspace{-0.5ex} Ultra High Energy (UHE) Tau neutrino may lead to a
very peculiar imprint in  future underground $Km^3$ detectors in
water and ice as well as in air: rarest secondary tau tracks and
decay which may exceed the muon ones. Indeed Bremsstrahlung  at
high energy lead to longer tracks for heavier leptons.  Radiation
lenght grows nearly with the square  of the lepton mass. Indeed
electrons are too light and their trace in matter (liquid or rock)
is negligible (tens of centimeters); muons are much better
observed, while tau are too short life time and too short range to
be found. However, because relativistic time expansion, UHE tau
traces in matter, above $10^{17}\;eV$, are relativistically
boosted overcoming  the corresponding muon tracks, already bounded
by bremsstrahlung logarithmic regime.  The tau   crossing for Kms
in water or ice may be confused with common muon tracks; their tau
decay may be misunderstood as  muonic catastrophic bremsstrahlung
interactions. To economize UHE tau discovery, we suggest  to look
the tau  decay in air into the deep valleys mountains, like
Canyons or deep in excavation mines where horizontal air showers
induce fluorescent or Cerenkov lights. The mountain valley width
screens from horizontal secondary muons. The valley  height
increases the solid angle view. The horizontal air Kms-size gap
offer  a strong discriminator to filter UHE muons against tau.
Tens event a year at PeV ( W resonance peak) energies in $Km^3$
excavation gap should be observable . Hunting air shower in the
night toward high mountains in Canyons or in a deep excavation may
be the best and cheapest way to discover UHE neutrinos , either
born by  electron anti neutrino scattering on electrons  at PeV
energies, or by direct tau neutrino possibly relic of muonic
flavour oscillation even at EeV energies.

\vspace{1ex}

\section{Introduction:}
Ultra high energy , UHE, neutrino astrophysics deals mainly with
muonic and electronic ones because they are the natural
secondaries of pion decays. Indeed  UHE tau and associated
neutrino tau are harder to be born by proton-proton  interactions
in active galactic nuclei (AGN).  Moreover their secondary charged
tau in the detectors are difficult to be noticed because of their
extremely brief life time $\simeq 3 * 10^{-13} sec$ and consequent
short tracks $5 \;mm \left(\frac{E}{100 GeV}\right)$. Nevertheless
these arguments hold only at "low" ($E_\nu \ll 10^5 GeV$)
energies. There are at least three valid argument that favor a
dominant key role of UHE and neutrino astrophysics contrary to
popular believes:\\ 1) The main stopping power for charged
leptons, the bremsstrahlung radiation and associated pair
production, is responsible for the (charged) lepton lengths. The
bremsstrahlung (and pair production) radiation length is
proportional (out of a logarithmic term) to the square of the
lepton mass. The heavier the mass the longer the lepton track
length. For this reason the longest tracks  should be the tau ones
$R_\tau$ as soon as the Lorentz boosted lifetimes and tracks,
growing linearly with energies, will reach the corresponding,
(bremsstrahlung  bounded by logarithmic growth) muon lengths
$R_\mu$. The peculiar phenomena (in the rock) occurs at energy $E
= 10^8 GeV.$ The longer tau traces makes UHE more detectable than
muon ones. The tau dominance occurs in a wide energy windows:
$10^{18}\;eV < E < 10^{22}\;eV$, within and above GZK cutoff
(Fargion et al.1999) \\ 2) On the other side the UHE source may be
naturally born by neutrino oscillation at widest neutrino mass
ranges, (because of the huge cosmic distances) and at the maximal
mixing angles, as strongly suggested by last Superkamiokande data.
\begin{equation}
L_{\nu_\mu \rightarrow \nu_\tau} = 1.23 \cdot 10^{13} cm
\left(\frac{E_\nu}{10^{17} eV}\right) \left(\frac{\triangle
m_{i\,j}}{eV}\right)^{-2} \ll L_{galaxy} \ll L_{cosmic}.
\end{equation}
3) While UHE tau are able to decay in air at bounded
\begin{equation}
R_{\tau_0} = 5 Km \left(\frac{E_\tau}{10^8 GeV} \right)
\end{equation}
distances in Earth, the corresponding muon tracks, being able to
travel above ten thousand Kms, fly longer with no decay above the
atmosphere. The difference may be revealed in $Km^3$ void traps
like deep canyons or valley.\\ 4) Anti neutrino electron
scattering on matter electrons in the W boson resonant peak, at
PeV energies, will produce the largest cross sections and also
secondary tau. Their signal must be observable even if no mixing
and no neutrino tau sources are available.\\ 5) Tau tracks,
because of their boosted radiation lengths, grow linearly with
energy (with respect to muon tracks) up to two order of magnitude
above muons ($ E > 4 \cdot 10^{10}\,GeV$) bounded by corresponding
radiative pair production length $R_{R\,\tau}$. However the
presence of weak interaction between UHE (as well as with matter
nuclei) grows, leading at $E = 4 \cdot 10^9\;GeV$, to a more
restrictive interaction length than radiative one $R_{W\,\tau}$,
for secondary tau as shown in figure below, reaching a peak in the
$R_{\tau}$ curve.\\ As shown in (Gandhi et al. 1998) above the PeV
energies expected neutrino might lead to few events in $Km^3$
detectors in most neutrino flux models. The relevant lepton tracks
in the detectors are the following (Fargion 1997,1999):
\begin{equation}
 R_{R_{\tau}} \cong 1033 \; Km \,
\left(\frac{\rho_r}{5}\right)^{-1} \, \left\{\, 1 \,+\,
\frac{\ln\left[\left(\frac{E_{\tau}}{10^8 \,
\mathrm{GeV}}\right)\left(\frac{E_{\tau}^{\min}}{10^4 \,
\mathrm{GeV}}\right)^{-1}\right]}{(\ln \, 10^4 )}\right\} \; .
\end{equation}
\begin{equation}
 R_{W_{\tau}} = \frac{1}{\sigma N_A \rho_r} \simeq
2.6\cdot 10^3 \, \mathrm{Km} \;
\left(\frac{\rho_r}{5}\right)^{-1}\, \left(\frac{E_{\tau}}{10^8\,
\mathrm{GeV}}\right)^{-0.363} \; .
\end{equation}
\begin{equation}
 R_{\tau} = \left( \frac{1}{R_{R_{\tau}}} +
\frac{1}{R_{\tau_o}} +\frac{1}{R_{W\tau}}\right)^{-1} \; .
\end{equation}
\begin{equation}
 R_{\mu} \left(E_\mu \gg 10^4\; GeV\right) \simeq 7.9 Km \; \left(\frac{\rho_r}{3}\right)
 \ln \left(\frac{E_\mu}{10^8\,GeV}\right).
\end{equation}
For more details see (Fargion 1997). The deep valley which we
consider as the ideal $Km^3$ detectors for horizontal tau air
showers are found in South Africa diamond mine excavations of
Kimberley. Besides the large mountain chain in Asia (Tibet),
Africa, South America (Andes) there are very good candidate large
and deep valley in Nevada, the Valley within White, Inyo and
Whitney Mountains, near the Death Valley. There are also wide
Canyons and one side sharp mountains to be considered. In Italy we
suggest the deep valley in the Alps, the Glacier du Miage at Mont
Blanc. We suggest also the nearby Mer de Glace and the Glacier
d'Argentier in the same mountain at French side. Also The
estimated event rate for a given surface detector are larger than
corresponding horizontal air shower rates because of the larger
density and volumes of the calorimeter mountains responsible for
the neutrino-nucleon interactions. \\ Even the (unexpected)
absence of any tau showers in air in new neutrino detectors will
imply puzzling constrains on SuperKamiokande data and
astrophysical neutrino models. In conclusion, we believe that in
future Km$^3$ telescope more surprises may (and must) come from
neutrino tau and tau signals: the first direct $\nu_{\tau}$
experimental evidence, its possible flavor mixing and the first
possible spectacular insight at highest energetic $(\geq 10^8\div
10^{11} \,GeV)$ neutrino astrophysical frontiers.
\vspace{1ex}
\section{Figures:}
\begin{figwindow}[1,r,%
{\mbox{\epsfig{file=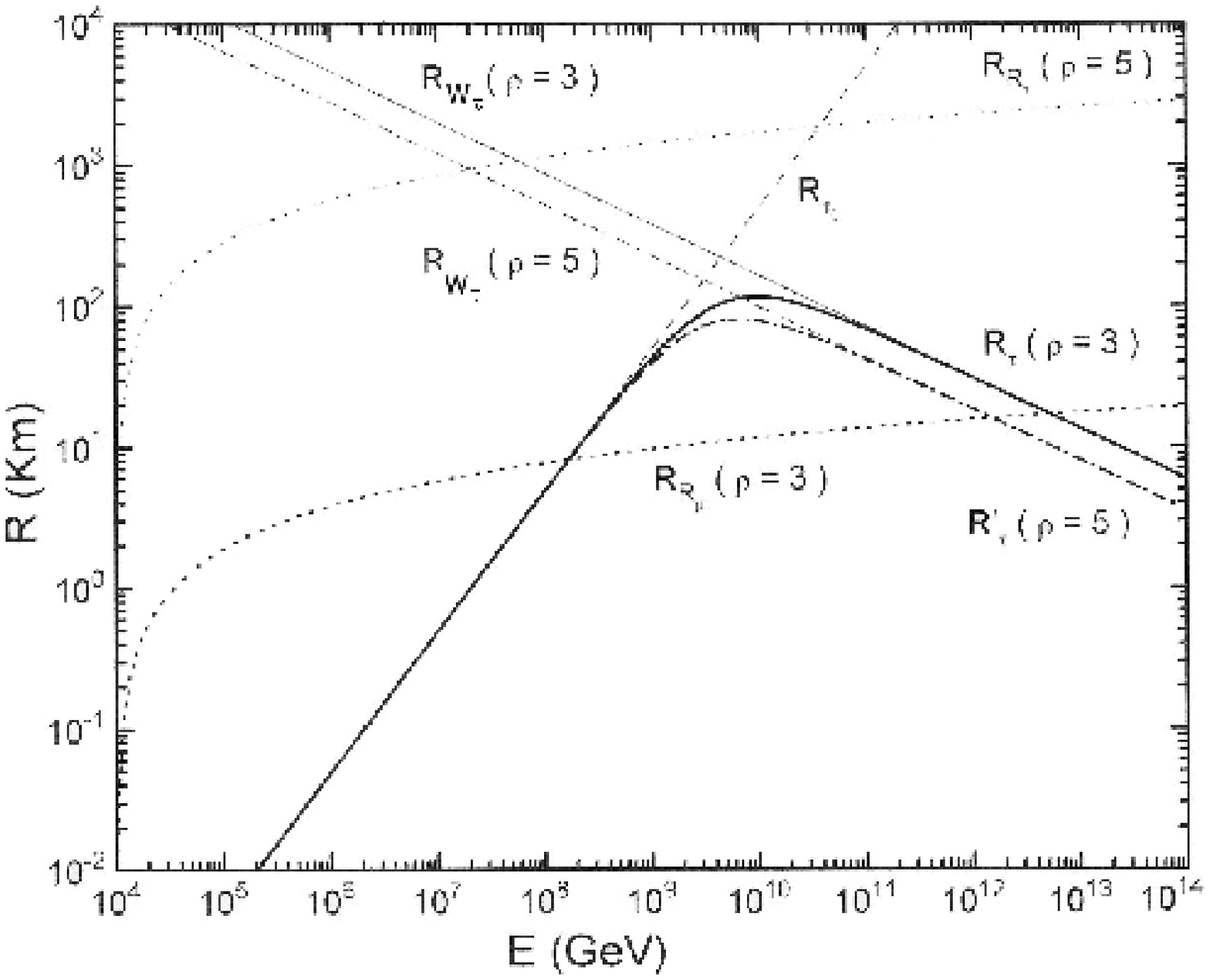}}},%
{}]
\end{figwindow}
%
%
\vspace{1ex}
\begin{center}
{\Large\bf References}
\end{center}
Fargion, D. 1997, Astro-ph 9704205, submitted final version to ApJ
1999\\ Fargion, D., Mele, B., Salis, A. 1999, ApJ 517, 725\\
Ghandi, R., Quigg, C., Reno M. H., Sarcevic I. 1998,
Hep-ph/9807264\\

\end{document}